\documentclass[12pt]{article}

\pdfoutput=1

\usepackage{graphics}
\usepackage{amssymb}
\usepackage{amsmath}
\usepackage{hyperref}
\usepackage{tikz}

\newcommand{\beq}{\begin{equation}}

\newcommand{\eeq}{\end{equation}}
\newcommand{\bea}{\begin{eqnarray}}
\newcommand{\eea}{\end{eqnarray}}

\begin{document}

\begin{center}
${}$\\
\vspace{100pt}
{ \Large \bf Measuring the Homogeneity (or Otherwise)\\ \vspace{10pt} of the Quantum Universe 
}

\vspace{46pt}

{\sl R.\ Loll}
and {\sl A.\ Silva}

\vspace{24pt}
{\footnotesize

Institute for Mathematics, Astrophysics and Particle Physics, Radboud University \\ 
Heyendaalseweg 135, 6525 AJ Nijmegen, The Netherlands.\\ 
\vspace{12pt}
{email: r.loll@science.ru.nl, agustin.silva@ru.nl}\\
}
\vspace{48pt}

\end{center}


\begin{center}
{\bf Abstract}
\end{center}

\noindent 
There are not many tools to quantitatively monitor the emergence of classical geometric features from 
a quantum spacetime, whose microscopic structure may be a highly quantum-fluctuating ``spacetime foam".
To improve this situation, 
we introduce new quantum observables that allow us to measure the absolute and relative homogeneity of geometric properties of a nonperturbative 
quantum universe, as function of a chosen averaging scale. 
This opens a new way to compare results obtained in full quantum gravity to  
descriptions of the early universe that assume homogeneity and isotropy at the outset.
Our construction is purely geometric and does not depend on a background metric.
We illustrate the viability of the quantum homogeneity measures by a nontrivial application to two-dimensional Lorentzian quantum
gravity formulated in terms of a path integral over Causal Dynamical Triangulations, and find some evidence of quantum inhomogeneity.

\vspace{12pt}
\noindent


\newpage

\section{Introduction}
\label{sec:intro}
Suppose we were given a quantum spacetime, obtained from a nonperturbative gravitational path integral or some other 
formulation of quantum gravity, and describing a (near-)Planckian regime. We will not assume a smooth
background structure or the availability of tensor calculus, but will rely on the presence of well-defined notions of distance and volume to
characterize the microscopic properties of the quantum spacetime. One question we are particularly interested in 
is to what extent a given quantum spacetime can be related to a description of the very early
quantum universe. For instance, can we show that familiar properties of classical models of the pre-inflationary 
universe ``emerge'' from it on sufficiently large scales, or that it contains the quantum-gravitational seeds of structure formation? 

In contrast to fully fledged quantum gravity, quantum cosmology
aims to capture the quantum dynamics of the universe as a whole in terms of a small number of global variables, tantamount to making strong symmetry
assumptions about the underlying spacetime \cite{bojowald2020}. 
In the most extreme case, when the spatial geometry is taken to be 
homogeneous and isotropic, the quantum dynamics of spacetime is reduced to the quantum mechanics of a single degree of freedom, 
the spatial volume (or scale factor).    
It would be interesting to understand whether some justification for such symmetry reductions can be derived from a theory of quantum gravity.

With this motivation in mind, we will below construct quantum observables that enable us to quantitatively assess the degree of
homogeneity of particular quasi-local properties of a quantum spacetime in a nonperturbative regime of full quantum gravity,  
focusing on the pure-gravity case. In addition to our theoretical considerations, we will present explicit computational results for a 
two-dimensional toy model of quantum gravity, to
illustrate some of the technical and conceptual issues in constructing measures of homogeneity.

A heuristic image often associated with quantum spacetime near the Planck scale is that of a ``foam'', invoking the possibility of a nontrivial
microstructure of spacetime, dominated by tiny wormholes or other nonclassical excitations \cite{wheeler1957} (see also \cite{carlip2022} for a recent review).
To ensure compatibility with the spacetimes of General Relativity, one then has to assume the existence of an averaging mechanism that 
produces macroscopically smooth spacetimes from microscopically ``foamy'' and possibly singular structures. In practice it is difficult to come up
with well-motivated quantum models of spacetime foam and even more challenging to demonstrate that any nontrivial microstructure ``averages out" on larger
scales. 

The aim of the present work is to develop concrete tools to investigate the degree of homogeneity of a quantum spacetime, 
whose local structure may be foam-like or carry some other form of quantum excitations.
As should be clear from our discussion above, the notions of scale and averaging will play important roles. 
Since by assumption we will work in a nonperturbative regime, we must make sure that any characterization of a
quantum spacetime is well defined operationally. This implies that we define and evaluate suitable diffeomorphism-invariant observables, 
and extract the physical properties of quantum spacetime from their expectation values. 

Identifying such observables in a nonperturbative context is not straightforward.\footnote{Further considerations about observables in 
nonperturbative quantum gravity can be found in \cite{questions2022}, Secs.\ Q28 and Q29. Nontrivial issues already appear at a perturbative level, 
see for example \cite{Donnelly2015,Frob2017} and references therein. } On the one hand, diffeomorphism invariance and the absence of a 
fixed, a priori background
structure imply that typical observables are nonlocal, for example, spacetime integrals of scalars or of more complicated
composite quantities. At the same time, observables will generally suffer from 
ultraviolet divergences, requiring a nontrivial regularization and renormalization. 
A nonperturbative path-integral approach to quantum gravity that has made significant advances in obtaining a quantitative grip on observables in a Planckian
regime is that of Causal Dynamical Triangulations, or CDT for short (see \cite{review1,review2} for reviews). A key result that has motivated the
current work is the emergence of an extended quantum spacetime, dynamically generated in 
four-dimensional CDT quantum gravity \cite{Ambjorn2004,Loll2007}. Several quantum observables have been investigated to understand its physical properties,
including its (average) Hausdorff and spectral dimension \cite{Ambjorn2005a,Ambjorn2005}, overall shape (volume profile) \cite{Ambjorn2007,Ambjorn2008}
and average scalar curvature \cite{Klitgaard2020}. Remarkably, their behaviour on sufficiently large scales has been found to be compatible
-- in the sense of expectation values and within the measuring accuracy of the numerical simulations -- with that of a classical de Sitter space in four
dimensions. This is intriguing because in standard cosmology the early universe is usually {\it assumed} to be de Sitter-like, whereas the cited quantum gravity
results have been derived dynamically and from first principles, without making any symmetry assumptions at the outset. 

These findings are highly nontrivial, but
it is important to understand that they do not imply that the geometric properties of this nonperturbative 
quantum spacetime are approximated in any {\it (quasi-)local} sense by those of a classical de Sitter space, let alone describable by a local 
metric tensor $g_{\mu\nu}(x)$.    
In fact, the typical diameter of the quantum spacetimes investigated in CDT so far is on the order of 15--20 Planck lengths \cite{review1}, and quantum
fluctuations of quasi-local quantities like the above-mentioned spectral and Hausdorff dimensions and quantum Ricci scalar are observed to be large. 
It would clearly be desirable to have well-defined observables which capture finer-grained information on the geometry of these quantum spacetimes, 
including their quantum fluctuations,
in order to relate and compare them to particular (semi-)classical cosmological
spacetimes. This is precisely what our quantum homogeneity observables are designed to do. 

We will study quantities associated with geodesic balls, which from a continuum point of view can be thought of as ball volume averages
of particular local scalar properties of spacetime. Note that because of discretization
effects, the analogues of local continuum quantities typically become quasi-local on the lattice, in the sense of being associated with a 
neighbourhood of a lattice vertex. Given such a quasi-local quantity on a spacetime lattice and a prescription for how to
average it over geodesic balls of size $\delta$, the associated quantum homogeneity measure will allow us to determine whether or not the quantity behaves 
approximately homogeneously on the scale $\delta$, in a way that will be spelled out in more detail below. 

The remainder of this paper is organized as follows. 
In Sec.\ \ref{sec:homo} we recall the classical notion of a spacetime isometry and discuss some of the difficulties one encounters when trying to translate
this concept to the nonperturbative quantum theory. We then introduce absolute and relative measures of homogeneity, at first in a classical context. 
They depend on a local or locally averaged geometric scalar quantity and on an averaging scale, and are global observables that characterize a given curved Riemannian manifold. We also introduce our choice of averaging regions, given in terms of a covering of the manifold by geodesic balls, whose radius sets the
averaging scale.  
In Sec.\ \ref{sec:impl}, we first describe the concrete implementation of the homogeneity measures in two-dimensional CDT quantum gravity, including a
prescription of how to cover CDT configurations by geodesic balls. We then discuss the results of our Monte Carlo measurements for the expectation
values of three different relative homogeneity measures, for the local deficit angle, the local coordination number, both of which are related to local curvature, 
and for the local Hausdorff dimension. All of them exhibit different behaviours as a function of the averaging scale. Sec.\ \ref{sec:disc} contains our conclusions and an outlook on future work.
In the appendix, we illustrate by an explicit example that our measures of homogeneity differ 
from those introduced in previous related work \cite{Cooperman2014}.

\section{Defining (in)homogeneity in quantum gravity}
\label{sec:homo}

In the context of classical general relativity, homogeneity refers to a spacetime which possesses a particular set of isometries. 
An isometry is a global continuous symmetry of a spacetime $(M,g_{\mu\nu})$ that leaves its metric structure invariant.\footnote{see 
e.g.\ \cite{Weinberg1972} for a concise introduction and discussion of isometries in classical gravity} 
The infinitesimal generator $K^\mu$ of an isometry is called a Killing vector field and satisfies the defining equation 
\begin{equation}
\nabla_\mu K_\nu+\nabla_\nu K_\mu \! =\! 0,
\label{killing}
\end{equation}
where $\nabla$ is the covariant derivative associated with
the metric $g_{\mu\nu}$. Note that the covariance of the Killing equation (\ref{killing}) implies that the presence of an isometry is independent of
the choice of coordinates. 

A general spacetime or solution to the Einstein equations has no Killing vectors at all.
A spacetime of dimension $D$ can have at most $D(D+1)/2$ independent Killing vectors, in which case it is called maximally symmetric.      
By definition, a homogeneous spacetime possesses $D$ independent Killing vectors, 
generating what can be thought of as the curved-space analogues of translations, 
i.e.\ a set of isometries that connects any two points of a (connected) spacetime manifold. A concept closely related to homogeneity is that of isotropy. 
A spacetime is isotropic in a point $p$ if there exist $D(D-1)/2$ independent Killing vectors vanishing at $p$, which can be thought of as the
generators of rotations or boosts about $p$. 
A spacetime is called isotropic if it is isotropic in every point. Isotropy implies homogeneity, but not the other way round.
For example, a homogeneous spacetime can have a distinguished direction from the point of view of its metric properties, thereby breaking isotropy.      

The task at hand is to determine whether and to what extent the concept of isometry can be made sense of in a nonperturbative regime of
quantum gravity where neither coordinate systems nor tensor calculus are available. We can immediately anticipate some difficulties when trying
to represent the {\it tensorial} character of eq.\ (\ref{killing}), or of the equivalent equation ${\cal L}_K g\! =\! 0$ (vanishing Lie derivative of the metric) 
in the quantum theory. The directional character of these relations is two-fold, both in the metric tensor itself and in the direction of the flow lines of 
the vector field $K$ along which the metric is Lie-dragged. 

For definiteness and later use, let us consider these issues in the framework of CDT, 
where quantum spacetime is obtained from a continuum limit of a regularized path integral, defined in terms of   
an ensemble of triangulated, piecewise flat metric manifolds \cite{review1}.\footnote{In technical terms, one aims to extract
a continuum theory by taking a scaling limit at one of the second-order phase transitions found in the lattice-regularized theory. 
In this sense, CDT plays the same role in gravity as lattice QCD does in nonabelian gauge theory, while taking into account background independence and
the dynamical character of spacetime geometry. }
Importantly, the individual regularized spacetime configurations are described 
in terms of their edge lengths and lattice connectivity, without recourse to coordinates \cite{Regge1961}. 
Nevertheless they possess well-defined, discretized notions of geodesic distance and volume, which may or may not scale canonically (i.e.\ according to their
naive classical dimensionality) in a Planckian regime in the continuum limit. 
Typical quantum observables that can be constructed from these elementary geometric ingredients are (simplicial versions of) 
scalar-valued spacetime integrals. 
Even if the quasi-local quantity that one integrates over has a directional character, like the quantum Ricci curvature \cite{Klitgaard2017,Klitgaard2018}
or a two-point function \cite{Ambjorn1996}, this direction-dependence will usually be lost 
upon integration. Likewise, without a differential manifold structure and tensor
calculus there are in general no obvious or useful analogues of classical concepts like parallel transport or the flow along the 
integral curves of a vector field, which would enable
us to compare direction-dependent metric properties at nearby points.\footnote{Ref.\ \cite{Brunekreef2020} illustrates some of the difficulties that arise when trying 
to define the concept of an approximate Killing vector on the piecewise flat geometries of the two-dimensional (C)DT path integral. We
will adopt a different strategy here.} 

It follows that any geometry-preserving symmetry we may define in a truly nonperturbative realm
will inevitably be a more basic and less stringent notion than that of an isometry of a (pseudo-)Riemannian geometry. 
Even then, one may expect on physical grounds that spacetime in a strongly quantum-fluctuating regime does not possess any symmetries of this kind. 
On Planckian scales, this may also be the case for the de Sitter-like quantum spacetime generated by the CDT path integral, which we already 
mentioned in the introduction.  
Even if this turns out to be true, it is conceivable that when examining the quantum spacetime through the lens of suitable averaged observables,
all Planckian inhomogeneities and anisotropies ``cancel out" above a certain averaging scale, such that  
in a statistical sense the system attains {\it approximate} symmetry properties which are compatible with those of a classical de Sitter universe.

Given our ambition to learn more about the possible quantum origin of our real universe, 
it would be very interesting if we could find evidence for such an emergence of symmetry from first principles {\it and} quantify deviations from it.
However, we should caution that the recovery of classical symmetries is by no means a foregone conclusion, neither in formulations based on
random geometry, like (Causal) Dynamical Triangulations, nor other nonperturbative models of quantum gravity, where a (background) spacetime is
not assumed at the outset. One known issue is that ensembles of microscopic building blocks have a tendency to exhibit collective behaviours which 
interfere with the emergence of large-scale four-dimensional geometry in the first place.\footnote{Two well-known effects, first discovered in 
Euclidean Dynamical Triangulations, are polymerization and crumpling, see e.g.\ \cite{Ambjorn1995,Catterall1995} 
as well as \cite{questions2022}, Sec.\ Q28, and references therein.} Also, depending on the specific quantity whose homogeneity
one investigates, the scale at which approximate homogeneity appears may lie outside the range of scales one has numerical access to.
  
The general prescription we will use to characterize the symmetry properties of a nonperturbative quantum spacetime implements a more
elementary idea than that of an isometry generated by a Killing vector field, namely that of ``a quantum spacetime looking
the same everywhere''. This is reminiscent of the so-called cosmological principle, which asserts that on suitably large scales the
spatial universe looks the same to all observers or, in more technical terms, that it is statistically homogeneous and isotropic, motivating the 
use of Friedmann-Lema\^{i}tre-Robertson-Walker (FLRW) models for their idealized, theoretical description.\footnote{see 
\cite{Abdalla2022,Aluri2022} for recent assessments of the observational evidence for the cosmological principle}

Due to the highly nonclassical nature of the geometries contributing to (C)DT path integrals it is not straightforward to define and parametrize (in terms of
angular variables, say) what one means by a local ``direction''.\footnote{see \cite{Klitgaardthesis}, Sec.\ 4.3, for an illustration in the context of 
quantum Ricci curvature in two-dimensional Euclidean quantum gravity in terms of Dynamical Triangulations (DT)} 
Even if one adopted a specific prescription, one would need a way to compare directions at different points, like parallel transport or distinguished 
global coordinates, which in a nonperturbative formulation are not readily available, as we explained earlier. For the time being we will therefore 
focus on scalar quantities, and will talk generically about their associated {\it measures of homogeneity}, which more precisely represent
statistical measures of homogeneity.

The main ingredient needed to specify such a measure of homogeneity on a given $D$-dimensional metric space $M$ of
positive definite signature is a scalar quantity $Q(x,\delta)$
associated with a geodesic ball $B_x (\delta)$ of radius $\delta$ based at the point $x$, which has the interpretation of a quantity $Q$ averaged
over the neighbourhood $B_x (\delta)$. 
In the simplest case, $Q(x,\delta)$ is the average of a local scalar $Q(x)$ over the ball $B_x (\delta )$,
which in a continuum notation reads
\begin{equation}
Q(x,\delta )\! =\! \frac{1}{ \mathit{vol}(B_x (\delta))}\int\limits_{B_x (\delta )}\!\! d^D y\, \sqrt{\det g}\; Q(y) ,\;\; \mathrm{with}\;\; 
\mathit{vol}(B_x (\delta ))\! =\!\!\!\int\limits_{B_x (\delta )}\!\! d^D y\, \sqrt{\det g}.
\label{olocal}
\end{equation}
The explicit examples we discuss in Sec.\ \ref{subsec:haus} below are all of this type.
We will comment in Sec.\ \ref{sec:disc} on the nonuniqueness of this choice of averaging.
It may happen that the quantities $Q(x)$ and $Q(x ,\delta)$ have an additional
dependence on some other scale $r$, which we will indicate by writing $Q(x; r)$ and $Q(x ,\delta; r)$.
The homogeneity or otherwise of a quantity $Q(x,\delta)$ across $M$, as a function of the averaging scale $\delta$, 
will be assessed by comparing it to its average $\bar{Q}( \delta )$ on $M$.
This requires a choice of how densely the points of $M$ will be sampled in the process. To obtain the maximal information about $Q( x ,\delta )$, we 
could compute its average by integrating over all $x\!\in\! M$\footnote{We will assume $M$ to be compact, which will always be the case
in our applications.} and compare this average pointwise to every $Q( x,\delta )$. However, to keep the computational effort manageable, and to limit
the amount of double-counting (i.e.\ how many times local information at a point $y$ appears in averaged quantities associated with different
geodesic balls $B_x(\delta )$), we will 
instead work with a discrete sample $S\!\subset\! M$ of points and their associated geodesic balls, and define the sample average $\bar{Q}[g](\delta )$ by
\begin{equation}
\bar{Q}[g](\delta )=\frac{1}{|S|}\, \sum_{x\in S} Q (x,\delta ).
\label{sample}
\end{equation}       
Here, $|S|$ denotes the number of points in $S$, and for later use
we have included the dependence on the metric $g$ (more precisely, on the diffeomorphism
equivalence class of $g$) explicitly in the notation.
The (nonunique) choice of a set $S$ should be such that $M$ is ``well covered'' by the associated
geodesic balls $\{ B_x (\delta )\, |\, x\!\in\! S\}$, in the sense that most points $y\!\in\! M$ should lie in exactly one ball, and the geometry of $M$ 
should be covered ``evenly",
without leaving large isolated regions uncovered. One could adopt a more stringent mathematical characterization of a such a covering or packing,   
but for our purposes the details of the prescription should be unimportant, as long as the linear size of $M$ is much larger than the ball radius $\delta$.
This should be verified in any explicit implementation. 
A convenient choices of $S$ for geometries in two-dimensional CDT quantum gravity will be discussed in Sec.\ \ref{sec:impl} below.

With these ingredients in hand, we can finally define the (absolute) homogeneity measure ${\cal H}_Q[g](\delta)$ of a scalar-valued quantity 
$Q(x, \delta )$ on a space with metric $g_{\mu\nu}$ by
\begin{equation}
{\cal H}_Q[g](\delta) := \sqrt{ \frac{1}{|S|\! -\! 1} \sum_{x\in S} \Big( Q( x, \delta) -\bar{Q}(\delta )\Big)^2 },
\label{homogabs}
\end{equation}
i.e.\ the square root of the unbiased sample variance of the set of densely sampled values $Q( x, \delta)$ with respect to the sample average, 
where we have suppressed the dependence on the set 
$S$ in the notation on the left-hand side. 
If the averaged quantity $Q$ is homogeneous at resolution $\delta$ on $(M,g_{\mu\nu})$, we have
${\cal H}_Q[g](\delta) \! =\! 0$, independent of the set $S$, while non-zero values signal the presence of inhomogeneities. 
The absolute homogeneity measure (\ref{homogabs}) is a natural choice in situations where absolute homogeneity
in the sense of ${\cal H}_Q[g](\delta)\! =\! 0$ can be attained exactly. However, this does not happen in the typical applications we have in mind,
where the value zero will at most be approached in a limit or in an approximate sense. In these cases we will often be interested in quantifying 
the inhomogeneities, that is, the degree to which homogeneity is violated, relative to some other scale relevant to the physical situation at hand. 
One natural choice is the (absolute value of the) $\delta$-dependent sample average $\bar{Q} (\delta )$ of the quantity $Q( x,\delta )$.
If it is nonvanishing, we can define a relative homogeneity 
measure ${\cal H}^{\it rel}_Q[g](\delta)$ by the dimensionless quotient
\begin{equation}
{\cal H}^{\it rel}_Q[g](\delta) := {\cal H}_Q[g](\delta)/|\bar{Q} [g](\delta )|, \;\;\;\;\;\;\;\;\;\;\;\;\;\;\;\;\; [\bar{Q}[g](\delta ) \not= 0].
\label{homogrel}
\end{equation}
Depending on the physical situation, other choices of reference scale may suggest themselves, which will then lead to alternative definitions of
relative homogeneity measures. In the remainder of this work, we will stick with the definition (\ref{homogrel}).
If the set $S$ is defined in geometric terms, without invoking any additional background structure that could bias the sampling,  
both of the measures (\ref{homogabs}) and (\ref{homogrel}) are diffeomorphism-invariant
classical observables, and suitable quantum implementations within the nonperturbative path integral are prime candidates for quantum observables. 
Explicit nontrivial examples will be discussed in Sec.\ \ref{sec:impl}. 

Since our homogeneity measures refer to specific quasi-local or averaged quantities, we can only claim that the underlying spacetime
is (approximately) homogeneous whenever (approximate) homogeneity has been shown for a complete set of such quantities. Classically, one might consider
using a complete set of local curvature invariants (whose number depends on the dimension), but these tend to not have well-defined analogues in
the nonperturbative quantum theory, where anyway few observables are available.  
When promoting the homogeneity measures to quantum observables below, we will therefore have little to say on how 
to define and construct a ``complete set'' of them. This being the case, one should be careful to refer to the homogeneity or otherwise 
{\it of the specific geometric quantity under consideration}, rather than of the quantum spacetime as such.

There are other instances in classical and quantum gravity where some form of averaging of geometry appears, 
for a variety of purposes. Perhaps closest in spirit to our quantum considerations is the question of how well
FLRW models describe the universe classically, and to what degree inhomogeneities of the real universe on smaller scales can lead to deviations from
the standard Friedmann dynamics on large scales. This ``averaging problem" or ``backreaction problem" in cosmology is the subject of ongoing research and 
faces multiple challenges, including how to average covariantly and beyond perturbation theory (see the reviews \cite{Ellis2011,Bolejko2016}  
and \cite{Schander2021} for the classical and quantum cases respectively).
Although we share some of these challenges, as already described earlier, our emphasis is different. We are not addressing questions of time evolution,
e.g.\ the (non-)commutativity of spatial averaging and solving equations of motion, but analyse the homogeneity properties of a given
``Riemannianized", nonperturbative (quantum) spacetime, without invoking any distinguished background geometry or presence of global symmetries 
at the outset.   

It is also worth pointing out that our averaging does not involve a coarse-graining of the underlying configuration space of the quantum theory, 
in the form of a decimation or ``integrating out" of the quantum (field) degrees of freedom associated with the microscopic details of geometry.  
Such a decimation,  accompanied by a suitable rescaling of coupling constants, typically appears in the renormalization group analysis of how a 
particular quantum field theory behaves as a function of scale, which for theories on Minkowski space is often identified with an ultraviolet momentum cutoff. 
Of course, adapting standard renormalization methodology to dynamical curved geometries beyond perturbation theory
is highly nontrivial both conceptually and technically. The difficulties in setting up a coarse-graining scheme for the simplicial manifolds used in (C)DT quantum gravity, analogous to spin blocking on a fixed
regular lattice, are well described in \cite{Henson2009} (see also references therein). 
Without relying on such a geometric coarse-graining scheme, implementations of renormalization group flows in CDT and open problems
have been discussed in \cite{Ambjorn2014,Ambjorn2020}. 
Considerable progress has also been made in continuum formulations of quantum gravity based on functional renormalization group methods \cite{Reuter2019}, 
but also here challenging issues remain \cite{Bonanno2020}. A difficulty common to all approaches concerns the definition and physical interpretation of 
the notion of ``scale" in a quantum theory of dynamical geometry.

Returning to the main subject of the current work, 
we will volume-average or ``coarse-grain" certain quasi-local geometric 
properties and their associated operators on a given quantum spacetime, rather than consider any renormalization group flows. 
In this way, we can probe its properties at various scales (``resolutions") 
of our choosing, which allows us to determine in quantitative terms whether a given property behaves (approximately) homogeneously in the range of physical scales considered. 
We assume that the renormalization of the underlying quantum gravity theory and its observables is well defined and under control, at least to the extent that
the quantum homogeneity observables we construct exist and are finite in the continuum limit. This is certainly the case for the homogeneity
observables for two-dimensional quantum gravity discussed in Sec.\ \ref{sec:impl} below.  

Lastly, let us comment on the relation of our construction with a previous attempt to introduce homogeneity measures in CDT quantum gravity
\cite{Cooperman2014}. Its author works with variances that superficially resemble our absolute homogeneity measure (\ref{homogabs}),
but without introducing a separate averaging scale (our ``$\delta$"). Instead, the notion of scale is associated with the scale-dependence of the
quantity whose variance is being investigated, which in our construction corresponds to the possible presence of another (linear) scale $r$, 
as mentioned just below eq.\ (\ref{olocal}).
There may of course be good reasons for doing so, but the resulting scale-dependent measures contain primarily 
information about the $r$-dependence of the underlying
quantity (and its variance) rather than about the degree of homogeneity as a function of averaging. As we will illustrate by
an explicit example in the Appendix, the dependences on $\delta$ and $r$ are in general different. 
One difficulty with interpreting $r$-dependent results is that the functional $r$-dependence of a generic quasi-local property of the quantum geometry
in a nonperturbative approach like CDT is not known a priori.\footnote{The two observables whose variance is studied in 
\cite{Cooperman2014}, the spectral dimension as a function of the
diffusion time $\sigma$ and the volume of a geodesic ball of radius $r$, are cases in point.} 
If it was, it could be taken into account to try and extract the associated inhomogeneities from the variances proposed in \cite{Cooperman2014}.   

By working with an explicit averaging scale $\delta$, related to the resolution of a geodesic ball covering, 
we address the question of how (in)homogeneously an averaged quantity 
$Q(x,\delta;r)$ behaves when $\delta$ is varied while $r$ is kept fixed. In an explicit example of this kind discussed in Sec.\ \ref{subsec:haus} below,
$Q(x,\delta;r)$ is the Hausdorff dimension extracted from measuring the volumes of spheres with a radius in the range $r\!\in\! [6,10]$,
averaged over a geodesic neighbourhood of radius $\delta$ based at the point $x$.

\section{Implementation and results in two dimensions}
\label{sec:impl}

Next, we will demonstrate that the homogeneity measures (\ref{homogabs}) and (\ref{homogrel}) can be implemented in 
a nontrivial model of nonperturbative quantum gravity, namely in the two-dimensional gravitational path integral formulated in terms of CDT,
which is an exactly soluble quantum gravity model with Lorentzian signature \cite{Ambjorn1998}.
It is inequivalent \cite{Ambjorn1999} to Liouville quantum gravity, which is based on a two-dimensional {\it Euclidean} path integral that
can be obtained from purely Euclidean Dynamical Triangulations \cite{David1993}.
Two-dimensional CDT quantum gravity is a well-studied toy model amenable to analytical methods 
(for a recent exposition, see \cite{Durhuus2022} and references therein),
and serves as a testing ground for new quantum gravity observables, most recently, the quantum Ricci curvature \cite{Brunekreef2021}. 
We will use it here to implement and test the new quantum homogeneity observables numerically. 

After Wick rotation, the CDT path integral takes the simple form
\begin{equation}
Z = \sum_{T} \frac{1}{C_T} \, {\rm e}^{-S [T]}, \quad \quad S [T] = \lambda\, N_2(T),
\label{2dpi}
\end{equation}
where the sum is taken over all geometrically distinct triangulations $T$ assembled from $N_2$ triangular Minkowskian building blocks, such
that $T$ has a well-defined causal structure \cite{Ambjorn2001a,review1}. 
Furthermore, $\lambda$ denotes the bare cosmological constant and $C_T$ the order of the automorphism group of the triangulation $T$.
We will adopt the standard choice of a $S^1\!\times\! S^1$-topology for
the triangulated spacetimes, with compact spatial slices and a cyclically identified time direction for computational convenience. 
As usual in CDT, there is a discrete proper-time parameter $t\!\in\! [0,1,2,\dots,t_{\rm tot}]$, with times 0 and $t_{\rm tot}$ identified, 
and where $\ell (t)\!\geq \! 3$ denotes the discrete length of the spatial universe at time $t$. 
Since the integral of the Ricci scalar in two dimensions is a topological invariant,  
we have dropped it from the Euclidean gravitational action $S_\lambda [T]$ in eq.\ (\ref{2dpi}), which therefore consists only of a cosmological-constant term. 

The expectation value $\langle {\cal O}\rangle$ of a geometric quantum observable ${\cal O}[T]$ in the ensemble of path integral configurations
is defined as
\begin{equation}
\langle {\cal O}\rangle =\frac{1}{Z}\ \sum_{T} \frac{1}{C_T} \, {\cal O}[T]\ {\rm e}^{-S [T]},
\label{expval}
\end{equation}
where ${\cal O}[T]$ denotes a suitable implementation on the piecewise flat space $T$ of a corresponding continuum quantity ${\cal O}[g]$.
We employ a Monte Carlo Markov chain algorithm to generate sequences of independent CDT geometries, allowing us to approximate 
the expectation values (\ref{expval}) by importance sampling of the full CDT ensemble.\footnote{We thank J.\ Brunekreef for sharing his CDT 
code.} As usual in CDT quantum gravity, we perform simulations with ensembles of fixed volume $N_2$ and extract the  
continuum limit $N_2\!\rightarrow\!\infty$ from a sequence of measurements on finite triangulations of increasing 
$N_2$, using finite-size scaling (see e.g.\ \cite{Newman1999}). 
We work with CDT lattices in the volume range $N_2\!\in\! [50.000,350.000]$ and for time extensions $t_{\rm tot}\!\in\! [62,200]$,
where as a result the average length $\bar{\ell}(t)$ of a spatial slice at integer time $t$ is of the order of 400--900.
During data collection, we have performed $N_{2}\!\times\! 1000$ attempted moves between measurements.
Considering that the equilibrium acceptance ratios are around $10\%$, every site of the geometry was on average updated roughly $100$ times between
consecutive measurements.

In the context of the lattice calculations, $x$ will denote a vertex of a triangulation $T$, and $\delta$ (and other measures of linear distance) 
will be given in terms of the link distance. The (geodesic)
link distance of two vertices $x$ and $y$ is defined as the number of links (edges) in the shortest path of contiguous links between $x$ and $y$.
Given a vertex $x$, the sphere ${\cal S}_x(\delta)$ of radius $\delta$ at $x$ is defined as the set of all vertices at link distance $\delta$ from $x$.
Similarly, the geodesic ball $B_x (\delta)$ of radius $\delta$ at $x$ -- for which we will use the same notation as in the continuum -- is given by the set of all vertices whose link distance to $x$ is smaller or equal to $\delta$. For a coarse-grained lattice quantity $Q(x,\delta )$, which is itself of the
form of a geodesic-ball average of a local scalar $Q(x)$, the analogue of the relation (\ref{olocal}) reads
\begin{equation}
Q(x,\delta )\! =\! \frac{1}{ \mathit{vol}(B_x (\delta))}\sum\limits_{y\in B_x (\delta )}\!\!  Q(y) ,\;\; \mathrm{with}\; \;
\mathit{vol}(B_x (\delta ))\! =\!\!\!\sum\limits_{y\in B_x (\delta )}\!\! 1.
\label{lattolocal}
\end{equation}

Before presenting our numerical results, we still need to specify how we choose the set $S$ of points 
that will serve as centres of the sphere covering in the lattice versions of
the homogeneity measures (\ref{homogabs}) and (\ref{homogrel}), namely
\begin{equation}
{\cal H}_Q[T](\delta)\! :=\! \sqrt{ \frac{1}{|S|-1} \sum_{x\in S} \Big( Q( x, \delta) -\bar{Q}[T](\delta )\Big)^2 },\;\; {\rm with}\;\;
\bar{Q}[T](\delta )=\frac{1}{|S|}\, \sum_{x\in S} Q (x,\delta ) ,
\label{latthomogabs}
\end{equation}
and 
\begin{equation}
{\cal H}^{\it rel}_Q[T](\delta)\! =\! {\cal H}_Q[T](\delta)/| \bar{Q}[T](\delta )|, \;\;\;\;\;\;\;\;\;\;\;\;\;\;\;\;\; [\bar{Q}[T](\delta ) \not= 0].
\label{latthomogrel}
\end{equation}

\begin{figure}[t]
\centerline{\scalebox{0.4}{\rotatebox{0}{\includegraphics{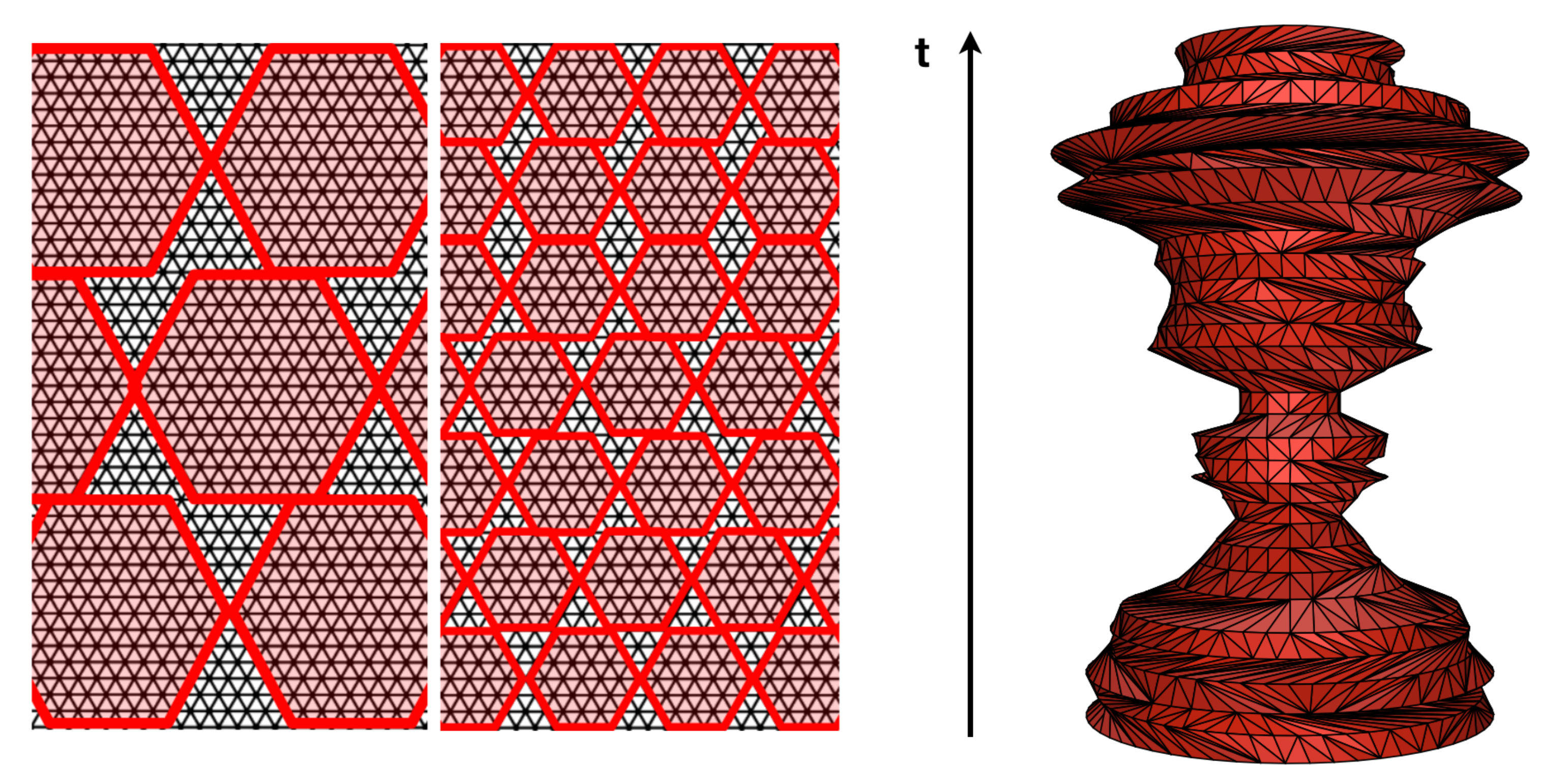}}}}
\caption{Pieces of an initial sphere packing as described in the text, for resolution $\delta\! =\! 6$ and $\delta\! =\! 3$ (left). 
For ease of illustration, the packing has been implemented on a regular, everywhere flat CDT configuration.
Generic CDT geometries carry local curvature and have spatial slices whose size $\ell (t)$ fluctuates strongly as a function of the time $t$,
as illustrated by the sample configuration on the right.}
\label{fig:spherepacking}
\end{figure}

To generate a vertex sampling $S$ of resolution $\delta$ on a given CDT configuration $T$, we start by constructing an (initial) sphere covering 
based on a vertex set $S^{\rm ini}$ of $T$. 
(This is not yet the final sphere covering that will be used to measure inhomogeneities.)
To obtain the sphere covering, we make use of the product structure of CDT configurations. 
We randomly pick a spatial slice $t_0$ and on it a vertex $x_0$, and construct a sphere ${\cal S}_{x_0}(\delta)$ of radius $\delta$ at $x_0$.  
Next, moving a distance $2\delta$ along vertices in the same slice, we pick a following vertex $x_1$ and construct its $\delta$-sphere,
which ``touches'' the first sphere
at time $t_0$. We continue moving along the slice in steps of length $2\delta$, constructing further spheres around vertices $x_2$, $x_3,\dots$, until no more 
spheres can be fitted without creating overlaps. 
Having thus constructed a sphere covering of a strip of time extension $\Delta t\! =\! 2\delta$, we repeat the same construction, using
as a starting vertex a randomly chosen vertex on the slice with label $t_0+2\delta$. After covering the corresponding strip $\Delta t\! =\! 2\delta$
by spheres, we progress in the time direction in steps of $2\delta$ until no more strips can be fitted in without creating nontrivial overlaps between
$\delta$-spheres (for illustration, see Fig.\ \ref{fig:spherepacking}, left). 

The sphere covering associated with the resulting finite set of vertices $S^{\rm ini}\! =\! \{x_0,x_1,x_2,\dots\}$
will in general have some gaps of linear size $\leq 2\delta$, but this should not matter much for triangulations
that are large compared to the scale $\delta$. The reason why we nevertheless do not use this sphere covering directly is a bias that
could potentially arise from the alignment of sphere centres along the selected spatial slices. Recall that the main dynamics of the CDT model is given by large
fluctuations of the spatial volume $\ell (t)$ as a function of time (Fig.\ \ref{fig:spherepacking}, right). If a quantity whose homogeneity is being assessed correlates
with the size of spatial slices, in particular when these are small, our initial selection $S^{\rm ini}$ of sphere centres may not be sufficiently random. The selection
principle we have adopted to avoid any such effects consists in picking one vertex from each of the spheres of the initial sphere covering and 
choose it as the centre of a sphere in a new sphere covering. The new set $S$ of circle centres is then spread roughly evenly over different spatial slices,
at the price of creating some overlaps between pairs of spheres, and therefore a slightly less complete covering of the total set of vertices in $T$.  
We have compared averaged observables for sphere coverings based on both prescriptions $S$ and $S^{\rm ini}$ for selected volumes, but have
not found any differences. All measurements reported below are based on choosing sets of type $S$. 

In what follows, we will evaluate the expectation value $\langle {\cal H}^{\it rel}_Q(\delta)\rangle$ 
of the {\it relative} homogeneity measure (\ref{latthomogrel}) of a quantity $Q$ in the quantum theory. 
Whether this or the absolute homogeneity measure are interesting and relevant depends on the context and application one has in mind.
We will use a straightforward 
implementation of the observable ${\cal O}[T]\! =\! {\cal H}^{\it rel}_Q[T](\delta)$ on each two-dimensional piecewise flat
CDT configuration $T$, where the set $S$ of vertices is selected as just described.
When measuring and interpreting homogeneity observables we must take into account the inherent limitations of the lattice regularization.
Firstly, to avoid finite-size effects due to the torus topology of the CDT configurations, we should only consider $\delta\!\ll\! L$, 
where $L$ is the linear size of a configuration $T$. In the case at hand this
is given by the time extension $t_{\rm tot}$ or the average spatial extension $\bar{\ell}\! =\! N_2/(2 t_{\rm tot})$, whichever is smaller, where
it should be kept in mind that the spatial length $\ell (t)$ fluctuates strongly as a function of $t$.\footnote{A quantitative analysis of the relation
between the average and the minimal value of $\ell$ for the same CDT ensemble was made in \cite{Brunekreef2021}.}
At any rate, nothing interesting can be learned when the resolution $\delta$ approaches the linear size of the spacetime,
since homogeneity is realized trivially there. 
In practice, we worked with an upper cutoff $\delta\! < \! t_{\rm tot}/4$. 
Secondly, results for small $\delta$ suffer from discretization 
artefacts and therefore should not be considered in any continuum interpretation. Previous investigations of geometric observables on
two-dimensional triangulations indicate that these discretization effects are definitely present on distance scales $\delta\!\leq\! 5$ in terms
of link units \cite{Klitgaard2017,Klitgaard2018}. In the explicit measurements discussed below, we find that this region is even larger, probably
due to the complex and somewhat extended nature of our construction.
Lastly, since we are interested in the homogeneity or otherwise of properties of the {\it continuum} theory, we study the expectation values of
the homogeneity measures for increasing lattice volumes and search for convergent behaviour as a function of a physical, rescaled  
averaging scale. 

The two homogeneity observables we will consider first involve averaged lattice quantities $Q(x,\delta)$ related to curvature. 
Both of them are geodesic-ball averages of a local quantity $Q(x)$, as in relation (\ref{lattolocal}) above. 
The local quantities are the coordination number $c(x)$, defined as the number of triangles meeting at the vertex $x$,
and the deficit angle $\varepsilon (x)$. Recall that a finite, piecewise flat triangulation in two dimensions 
carries intrinsic curvature, which is located in a singular way at its vertices in the form of deficit angles. For a given vertex $x$, the
deficit angle $\varepsilon$ is defined as $\varepsilon\! =\! 2\pi\! -\!\sum_i \gamma_i$, the difference of $2\pi$ and the sum of the angles
$\gamma_i$, where $\gamma_i$ is the angle contributed by the $i^{\rm th}$ triangle sharing the vertex $x$. 
Since the Wick-rotated CDT configurations are built from equilateral triangles, the deficit angle can be expressed in terms of the coordination 
number according to $\varepsilon (x)\! =\! 2\pi-c(x)\pi/3$. The deficit angle is a direct measure of the Gaussian curvature $K$ associated with the point $x$,
with $c(x)\! =\! 6$ characterizing a locally flat triangulation. Note that by virtue of the Gauss-Bonnet theorem and the toroidal topology of
the CDT configurations we have 
\begin{equation}
\sum_{x\in T} c(x)\! =\! 6 N_0,
\label{gb}
\end{equation}
where $N_0$ denotes the number of vertices in the triangulation $T$. 

In nonperturbative approaches like CDT which use simplicial manifolds as a regulator,
defining curvature in terms of deficit angles is problematic in dimension $D\! >\! 2$ because of ultraviolet divergences in the continuum limit 
(see e.g.\ the discussion in \cite{review2}). This has motivated the introduction of an alternative notion dubbed the quantum Ricci curvature \cite{Klitgaard2017,Klitgaard2018}. Since this problem is absent in two dimensions and a homogeneity measure based on
the quantum Ricci curvature is computationally expensive, we will investigate here homogeneity measures based on the deficit angle and
the coordination number.  

\begin{figure}[t]
\centerline{\scalebox{0.35}{\rotatebox{0}{\includegraphics{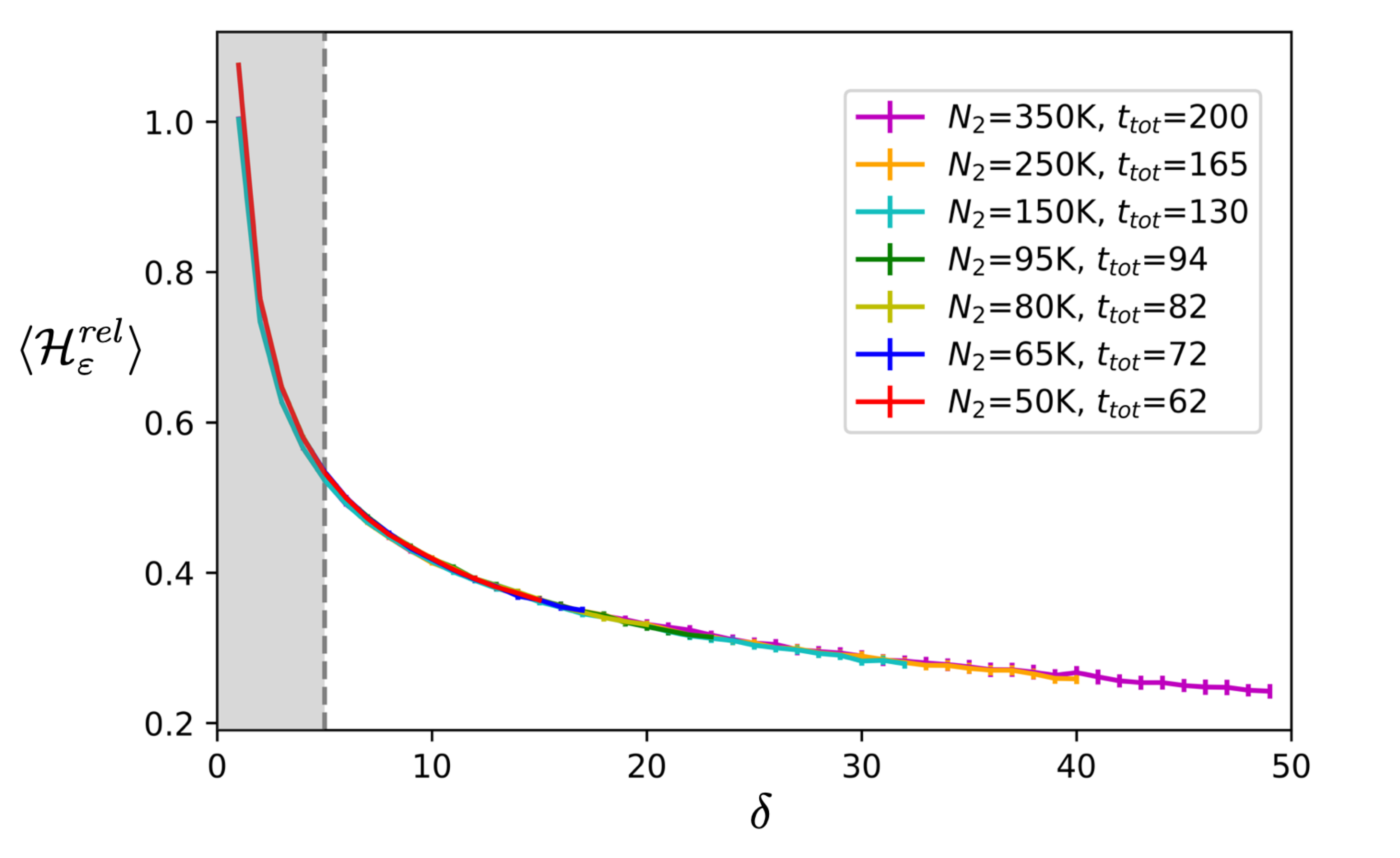}}}}
\caption{Expectation value $\langle {\cal H}^{\it rel}_\varepsilon(\delta)\rangle$ of the relative homogeneity measure associated with the averaged 
deficit angle $\varepsilon$, as a function of the averaging scale $\delta$ in lattice units, 
for volumes in the range $N_2\!\in\! [50k,350k]$ and time extensions $t_{\rm tot}\!\in\! [62,200]$, including statistical error bars. 
The shaded area $\delta\!\leq\! 5$ indicates the region dominated by discretization artefacts.}
\label{fig:deficit}
\end{figure}

\subsection{Homogeneity of the deficit angle}
\label{subsec:def}

We start by considering the deficit angle, $Q(x)\! =\! \varepsilon (x)$, which is a direct measure of the local curvature. 
For each fixed scale $\delta\!\in\! [1,49]$ we computed for each randomly sampled configuration $T$ of a CDT ensemble 
at fixed volume $N_2$ and 
time extension $t_{\rm tot}$ the averaged deficit angle $\varepsilon (x,\delta)$ for geodesic balls of radius $\delta$ associated with a sphere covering.  
The expectation value $\langle {\cal H}^{\it rel}_\varepsilon(\delta)\rangle$ in the quantum ensemble for a range of 
different volumes and as a function of the link distance is shown in Fig.\ \ref{fig:deficit}.
Data points at integer values of $\delta$ have been joined up by straight line elements to guide the eye. 
Note that the curves end at different values of $\delta$, due to our choice $\delta\!<\! t_{\rm tot}/4$. 
The initial region $\delta\!\leq\! 5$ is expected to be dominated by discretization artefacts. Our primary interest are the data points with $\delta\! >\! 5$,
which within error bars are seen to agree well for different volumes $N_2$ within their common $\delta$-ranges. 
For all lattice volumes, the homogeneity measure of the deficit angle is seen to decrease monotonically over the observed range, 
corresponding to an increase in homogeneity as a function of the averaging scale $\delta$, as one may have expected.  
However, since the relative homogeneity does not even fall to 24\% (its value at $\delta\! =\! 49$ is 0.242), this particular observable 
does not behave approximately homogeneously in the observed $\delta$-range. 

\begin{figure}[t]
\centerline{\scalebox{0.42}{\rotatebox{0}{\includegraphics{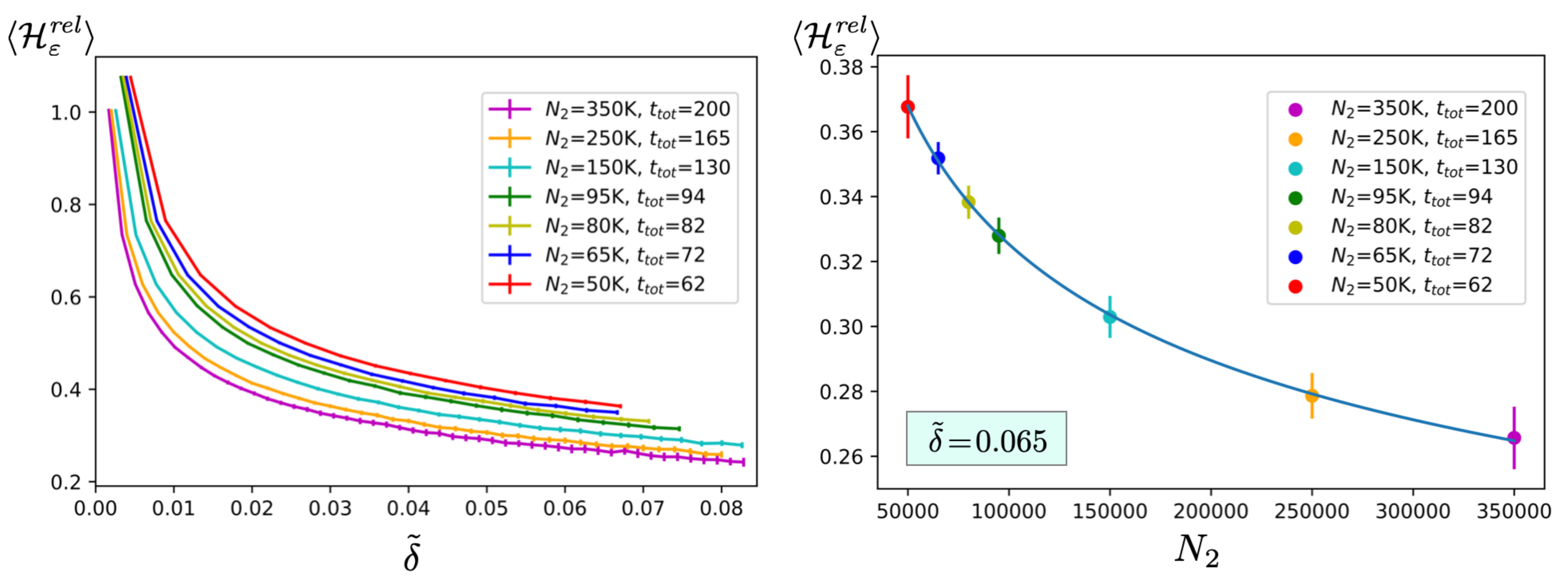}}}}
\caption{Expectation value $\langle {\cal H}^{\it rel}_\varepsilon \rangle$ of the relative homogeneity measure associated with the averaged 
deficit angle $\varepsilon$, as a function of the rescaled averaging scale $\tilde{\delta}$ (left). Expectation value $\langle {\cal H}^{\it rel}_\varepsilon \rangle$ at fixed $\tilde{\delta}\! =\! 0.065$, as a function of $N_2\!\in\! [50k,350k]$, together with a curve fit (right).}
\label{fig:deficit_scaled}
\end{figure}

To understand its behaviour in the continuum limit $N_2\!\rightarrow\!\infty$, we performed a finite-size scaling analysis by rescaling the lattice 
distances $\delta$ by the square root of the volume (Fig.\ \ref{fig:deficit_scaled}, left), and plotted the data as a function of the rescaled,
physical distance $\tilde{\delta}\! :=\!\delta /\sqrt{N_2}$.\footnote{We assume that the physical two-volume $V_2\! =\! a^2 N_2$ is kept fixed
as we scale $N_2\!\rightarrow\! \infty$, where $a$ denotes the dimensionful length of a lattice edge, such that $a\!\propto\! 1/\sqrt{N_2}$.}
The lack of overlap of the curves illustrates that the observable 
${\cal H}^{\it rel}_\varepsilon$ is subject to strong short-distance lattice artefacts, extending also beyond $\delta\! =\! 5$. 
Nevertheless, the presence of finite-size scaling can be exhibited by comparing the curves for sufficiently large $\delta$ and fixed 
physical distance $\tilde\delta$.   
When comparing data points at $\tilde{\delta}\! =\! 0.065$, at the end of the range for which
we have measurements from all volumes, we observe signs of convergence as a function of the volume $N_2$, as illustrated in Fig.\ \ref{fig:deficit_scaled}, right.
The interesting question is whether the homogeneity measure for sufficiently small $\tilde{\delta}\!\ll\! \tilde{L}$, where $\tilde L$ denotes the 
(rescaled) linear size of the spacetime, converges to a constant function in the continuum limit, 
and what the value of this constant is. We have fitted the seven data points at $\tilde{\delta}\! =\! 0.065$ to functions 
$F (N_2)=c_0+c_1 N_2^{-\nu}$, for selected $\nu$ with $0\! <\! \nu\!\leq 1$. The fit shown is for $F(N_2)\! \approx \! 0.100+4.00 N_2^{-1/4}$. The fit for
$\nu\! =\! 1/2$ is of comparable quality and yields $c_0\!\approx\! 0.205$. Both of these would imply significant levels of inhomogeneity (of 10\% and 20\%) 
in the continuum limit. However, it should be kept in mind that the limited volume range of our data set does not allow us to discriminate between
these different fall-off behaviours or to verify our simple ansatz for $F(N_2)$.

We also do not have stringent theoretical arguments for how $\langle {\cal H}^{\it rel}_\varepsilon(\delta)\rangle$ should behave in the limit. 
Regarding the denominator of this expression according to (\ref{homogrel}),
we have measured how the geodesic-ball average of the deficit angle behaves for $\delta\!\leq\! 100$,
and found excellent agreement with a scaling $\propto\! 1/\delta$ toward zero from below. 
On the other hand, {\it if} the original distribution of the values $\varepsilon (x)$ on a given configuration $T$
could be treated as independent random events (which is strictly speaking not true, but may still be a good approximation), 
it would follow from the central limit theorem that samples of size $n$ approach a normal distribution with standard deviation $\propto\! 1/\sqrt{n}$.
In the case at hand this would translate to a behaviour $\propto\! 1/\delta$, since the Hausdorff dimension of CDT configurations in the continuum limit is known 
to be $d_H\! =\! 2$ \cite{Ambjorn1998,Durhuus2009}. If this scaling scenario for the standard deviation
was realized, it would mean that the relative homogeneity measure 
$\langle {\cal H}^{\it rel}_\varepsilon(\delta)\rangle$ for large $\delta$ approaches a constant, which is consistent with what we found in our measurements. However, without a more detailed understanding of the behaviour of the variance for large $\delta$, we do not know which value (including zero) this
constant may have. 

\begin{figure}[t]
\centerline{\scalebox{0.35}{\rotatebox{0}{\includegraphics{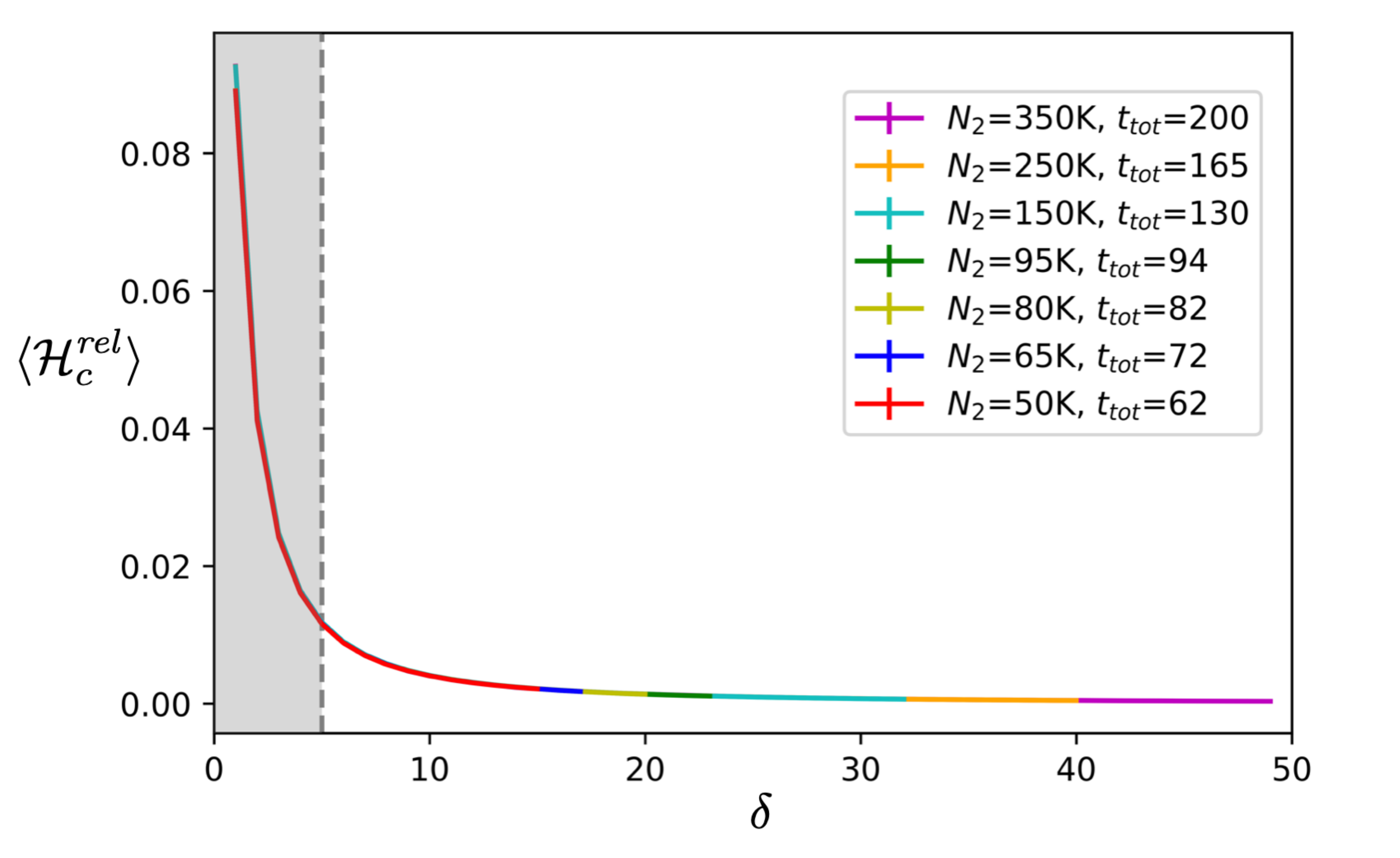}}}}
\caption{Expectation value $\langle {\cal H}^{\it rel}_c(\delta)\rangle$ of the relative homogeneity measure associated with the averaged 
coordination number $c$, as a function of the averaging scale $\delta$ in lattice units, 
for volumes in the range $N_2\!\in\! [50k,350k]$ and time extensions $t_{\rm tot}\!\in\! [62,200]$. Error bars are too small to be shown.
The shaded area $\delta\!\leq\! 5$ indicates the region dominated by discretization artefacts.}
\label{fig:coordination}
\end{figure}

\subsection{Homogeneity of the coordination number}
\label{subsec:coord}

\begin{figure}[t]
\centerline{\scalebox{0.42}{\rotatebox{0}{\includegraphics{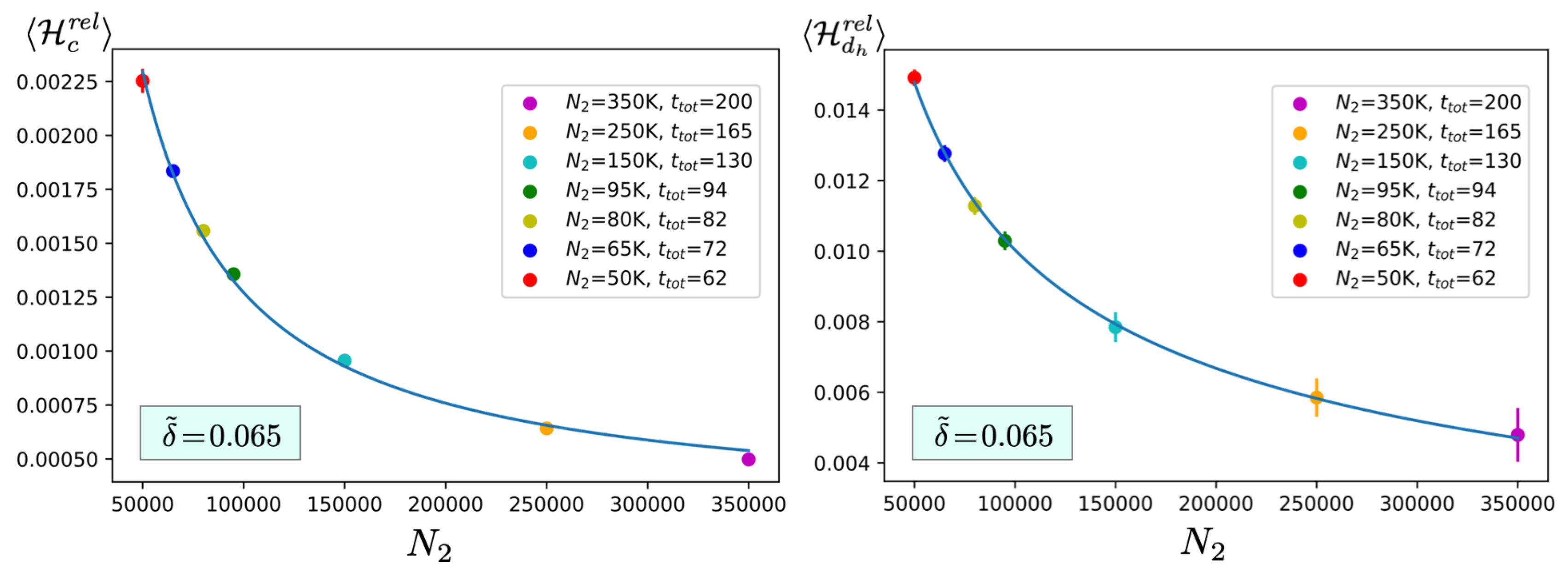}}}}
\caption{Expectation value $\langle {\cal H}^{\it rel} \rangle$ of the relative homogeneity measures associated with the averaged 
coordination number $c$ (left) and the averaged 
Hausdorff dimension $d_h$ (right) at fixed $\tilde{\delta}\! =\! 0.065$, as a function of $N_2\!\in\! [50k,350k]$, together with curve fits.}
\label{fig:coord_hausdorff_scaled}
\end{figure}

Looking instead at the relative homogeneity measure of the coordination number, $Q(x)\! =\! c(x)$, avoids the potential problem of dividing by an average
that is close to zero, since the geodesic-ball average of $c(x)$ will approach 6 for large $\delta$, by virtue of the global constraint (\ref{gb}).  
Repeating the same analysis as before yields the expectation values shown in Fig.\ \ref{fig:coordination}. Their absolute scale is 
of course smaller than that associated with the deficit angle of Fig.\ \ref{fig:deficit}, since we are dividing by a larger average. 
From the point of view of continuum physics, this absolute scale is really immaterial, since the coordination number as such does not have a 
continuum counterpart. Let us nevertheless proceed and accept the geodesic-ball average of $c(x)$ as an appropriate reference scale for use
in the homogeneity measure. Since this average is well approximated by a nonvanishing constant $\approx\! 6$, 
what we are looking at here is -- up to this overall constant -- the absolute homogeneity measure (\ref{latthomogabs}) 
of the coordination number. Fig.\ \ref{fig:coordination} shows that there is a very close overlap between the data for different volumes, and a rather
fast convergence toward a small expectation value as $\delta$ increases. We do not show the curves as a function of the rescaled distance $\tilde\delta$,
which look similar to those for the deficit angle, with slightly less severe discretization effects. More interesting is the comparison of 
the expectation values at fixed $\tilde{\delta}\! =\! 0.065$ (Fig.\ \ref{fig:coord_hausdorff_scaled}, left), 
which indicates convergence to a small value near zero. 
A fit to $F(N_2)\! \approx\! 2.46\! *\! 10^{-4}\! +\! 103 N_2^{-1}$ works well in this data range, subject to the same provisos as our earlier discussion
of the deficit angle. We thus conclude from our measurements that in the
$\delta$-range considered this particular observable attains a very high degree of relative homogeneity, in the sense of expectation values. 

A lesson we have learned so far is that the behaviour of the relative homogeneity depends on the quantity whose homogeneity is being assessed.
It is ultimately a physics question which quantities can be meaningfully considered from a continuum point of view. Computational feasibility is also important,
because of the dense sampling we perform of individual path integral configurations.
A tentative interpretation for the behaviour of the expectation value $\langle {\cal H}^{\it rel}_c(\delta)\rangle$ associated with the 
coordination number is that the only inhomogeneities observed are confined to a region near the cut-off scale (which anyway we
discard because of discretization artefacts), outside of which
the homogeneity is almost perfect. This is the behaviour one would generically expect from any local, randomly distributed lattice property.
In the case at hand, the homogeneity of the coordination number should probably be seen as a consequence of the unphysical nature of this local
lattice property, rather than an indication that the quantum geometry of two-dimensional CDT is homogeneous. 
At any rate, we have already seen evidence that the behaviour of the relative homogeneity measure $\langle {\cal H}^{\it rel}_\varepsilon(\delta)\rangle$ 
associated with the deficit angle $\varepsilon$, a quantity more closely related to curvature, may be less trivial. 
Our measurements on lattice volumes up to 350$k$ are compatible with a behaviour where 
the size of the fluctuations stays large compared to the mean value $\bar{\varepsilon}$, although larger lattices will be needed to corroborate this.

\subsection{Homogeneity of the Hausdorff dimension}
\label{subsec:haus}

The next quantity whose homogeneity measure we will consider is not related to curvature and is given by 
the (local) Hausdorff dimension $d_h$. By definition, $d_h$ characterizes
the growth of small geodesic spheres ${\cal S}_x(r)$ as a function of their radius $r$. 
In two-dimensional CDT quantum gravity, the Hausdorff dimension is known to be $d_h\! =\! 2$ analytically \cite{Ambjorn1998,Durhuus2009},
and has also been measured previously in numerical simulations \cite{Ambjorn1999a,Loll2015}. 
On a given triangulation $T$ of the CDT ensemble, the Hausdorff dimension 
$d_h(x)$ at a vertex $x\!\in\! T$ is extracted from the leading power law behaviour in $r$ of the sphere volumes
${\rm vol}({\cal S}_x(r))$, where
${\rm vol}({\cal S}_x(r))$ counts the number of vertices at link distance $r$ from $x$. We have used $r\!\in\! [6,10]$ as a fitting range for the radius, 
just beyond the region of lattice artefacts. Our fitting function was of the form
\begin{equation}
{\rm vol}({\cal S}_x(r))=c(x)\,  r^{d_h(x)-1},
\label{sfit}
\end{equation}
for a local multiplicative constant $c$.
It should be kept in mind that our primary aim is to establish homogeneity or otherwise of the {\it averaged} Hausdorff dimension, rather than
a precision measurement of the Hausdorff dimension per se. Due to the rather complex determination of $d_h(x)$, which involves geometric data from 
an entire neighbourhood of radius 10 around $x$, it is rather costly computationally to
extract the corresponding averaged quantities $d_h(x,\delta)$
according to the prescription (\ref{lattolocal}), especially for large geodesic balls ${\cal B}_x(\delta)$. 
To keep the computational effort manageable, we opted for the relatively short fitting interval [6,10] and the simple fit function (\ref{sfit}). 
We did not include an additive shift of $r$ in the fit, which is sometimes employed for optimization, after running some checks to make sure that
the effect on our results is small.    

Another feature worth noting is that because of the quasi-local nature of $d_h(x)$,
the geodesic-ball averaged quantities $d_h(x,\delta)$ depend also on the geometry {\it beyond} $\delta$,
more precisely, on data up to distance $\delta\! +\! 10$ from the central vertex $x$. This should be contrasted with the deficit angle
and coordination number, whose homogeneity we discussed above, which are genuinely local lattice quantities.
The quasi-local character of the Hausdorff dimension $d_h(x)$ is a characteristic feature of quantities with a more direct
continuum interpretation; other examples are the spectral dimension and the quantum Ricci curvature. 
While the quasi-local nature should make little difference in any continuum limit $N_2\!\rightarrow\!\infty$, 
how it affects our considerations of lattice versus physical effects at finite volume $N_2$ is a priori less clear.  
A possible effect of the quasi-locality could be that the curves of the homogeneity measures are shifted to some larger,
``effective'' $\delta$, compared to those for more local quantities. On the other hand, since the averaging over geodesic balls
now effectively contains some overlap between neighbouring balls, this may lead to an increase in the apparent homogeneity.  
 
\begin{figure}[t]
\centerline{\scalebox{0.35}{\rotatebox{0}{\includegraphics{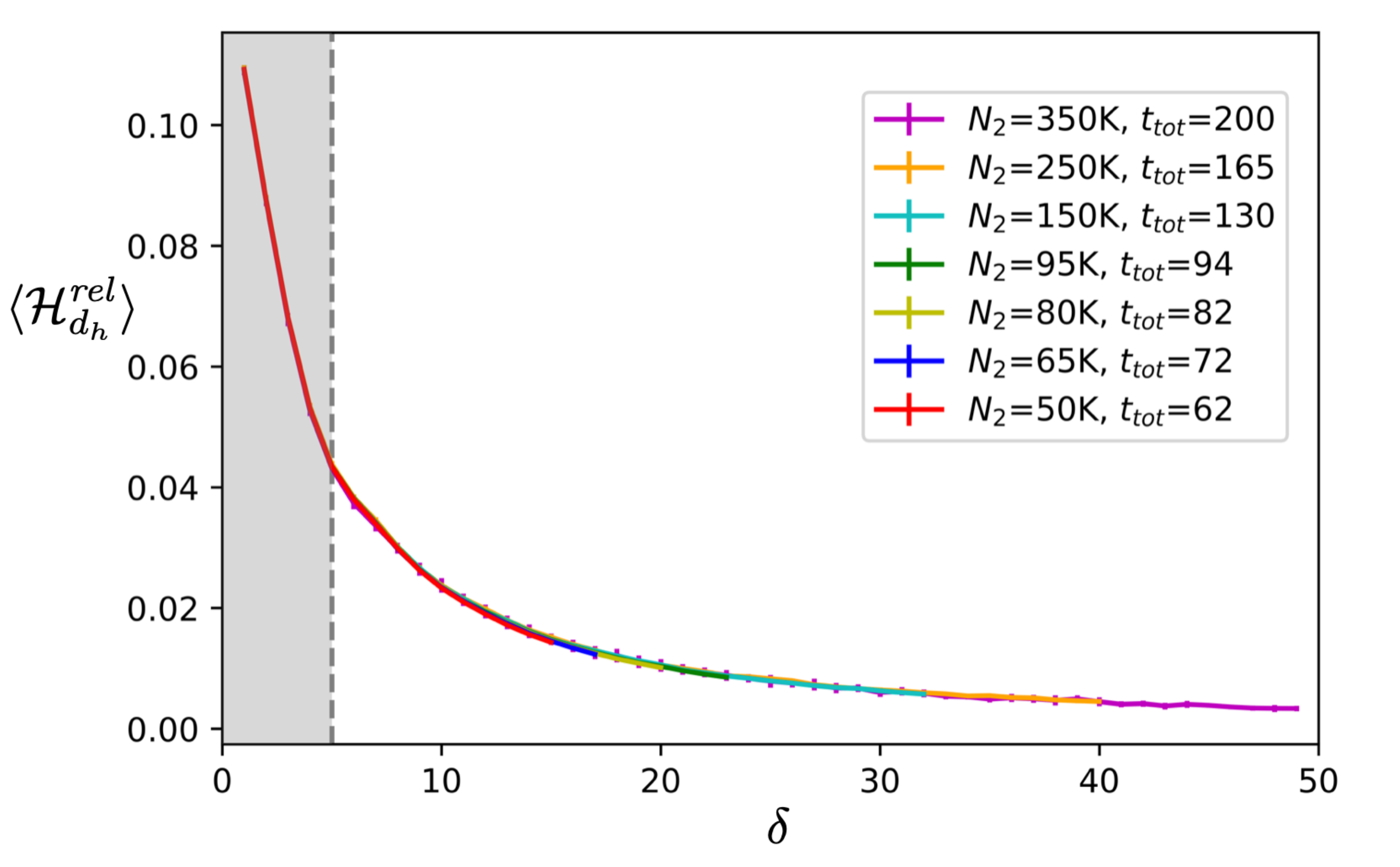}}}}
\caption{Expectation value $\langle {\cal H}^{\it rel}_{d_h}\rangle$ of the relative homogeneity measure associated with the averaged Hausdorff dimension $d_h(\delta)$, as a function of the averaging scale $\delta$ in lattice units,
for volumes in the range $N_2\!\in\! [50k,350k]$ and time extensions $t_{\rm tot}\!\in\! [62,200]$. The shaded area $\delta\!\leq\! 5$ 
indicates the region dominated by discretization artefacts. }
\label{fig:hausdorff}
\end{figure}

Our measurement data for the expectation value of the relative homogeneity measure are shown in Fig.\ \ref{fig:hausdorff}.
We again observe a monotonic decrease as a function of the averaging scale $\delta$, like for the measures we investigated earlier.
Since the Hausdorff dimension is a genuinely physical quantity, the data can perhaps best be compared to those of the homogeneity measure 
of the deficit angle. Compared to the latter, they indicate both a much greater degree of homogeneity and a faster fall-off for increasing
$\delta$ in the case of the Hausdorff dimension. The spread of the curves as a function of the rescaled length scale 
$\tilde{\delta}$ is similar to that shown for the deficit angle in Fig.\ \ref{fig:deficit_scaled}, left, but even bigger, 
indicating that the $\tilde{\delta}$-region with appreciable discretization effects is even larger. This is
compatible with the expectation that the quasi-locality of the Hausdorff dimension leads to a larger ``effective'' $\delta$ than in the case of the
more local deficit angle. The scaling with volume of the data at $\tilde{\delta}\! =\! 0.065$ (Fig.\ \ref{fig:coord_hausdorff_scaled}, right) is
within error bars compatible with an asymptotic approach to zero, using a fit with $\nu\! =\! 1/2$, which leads to
$F(N_2)\!\approx\! -1.42\! *\! 10^{-3}\!+\! 3.62 N_2^{-1/2}$. Also the choice $\nu\! =\! 1/4$ gives a fit of decent quality and yields $c_0\!\approx\! -0.012$. 
The difference between these $c_0$-values and the fact that both are negative illustrates the presence of systematic errors significantly larger than the
statistical error bars, which are on the order of only 2\%. We conclude that the Hausdorff dimension shows a highly homogeneous behaviour on CDT geometries, 
as quantified by the expectation value of our relative homogeneity measure (\ref{latthomogrel}). Taking into account the uncertainties present 
the outcome appears to be compatible with $\langle {\cal H}^{\it rel}_{d_h} \rangle (\tilde{\delta})\! =\! 0$ in the continuum limit.  

\subsection{Collecting results}
\label{subsec:collect}

Let us summarize what we have learned from implementing the relative quantum homogeneity measures in two-dimensional CDT
quantum gravity. We performed Monte Carlo simulations for three different quantities and for finite system sizes up to $N_2\! =\! 350.000$ 
triangular building blocks. We have focused on intermediate averaging scales $\delta$ above a lower cutoff $\delta\! =\! 5$ and much smaller than the total
linear system size, where a subsequent finite-size scaling analysis showed that the region where the homogeneity observables are affected by discretization 
artefacts extends beyond $\delta\! =\! 5$.     
Taking this into account when interpreting the data, we can draw some interesting conclusions.

All three homogeneity measures are monotonically decreasing in $\delta$, but the details and their interpretation are different.
The relative homogeneity measure of the coordination number $c$ falls off rapidly in $\delta$ to a value compatible with zero, and for
fixed physical scale $\tilde{\delta}$ approaches this limit with a rate roughly proportional to $1/N_2$. However, we attributed this
``perfect" statistically homogeneous behaviour to the unphysical nature of the coordination number, which does not have a direct continuum 
counterpart, rather than to the homogeneity of the underlying quantum geometry. The latter conclusion was supported by the fact that the
relative homogeneity measure of the deficit angle $\varepsilon$, which does have a direct correspondence with Gaussian curvature, shows a
different behaviour, with a slower fall-off in $\delta$. More importantly, our best fits from a finite-size scaling suggest an approach
proportional to a fractional inverse power $N^{-\nu}$, with $\nu\!\sim\! 1/4 ...1/2$, to a {\it nonvanishing} expectation value 
$\langle {\cal H}^{\it rel}_\varepsilon \rangle$ on the order of 10-20\%, indicating a residual statistical inhomogeneity in the continuum limit. 
Lastly, the relative homogeneity measure of the Hausdorff dimension $d_h$ behaves rather similar to that of the deficit angle, including the rate of
convergence to its continuum value. However, unlike what happens in the case of the deficit angle, 
this continuum value is zero within measuring accuracy. Because of the physical nature of
the Hausdorff dimension, we interpret this as a genuine continuum result. The fact that the dimension is more homogeneous than the
curvature may be attributable to the fact that the former is a more robust ``pregeometric" property, while the inhomogeneity of the curvature
may be a consequence of the pure quantum nature of two-dimensional quantum gravity, which does not have a classical limit. In particular,
there is no nontrivial classical solution that the quantum geometry could converge to in an $\hbar\!\rightarrow\! 0$ limit.

\section{Discussion and outlook}
\label{sec:disc}

Motivated by the search for quantitative measures of the degree of homogeneity of a quantum geometry at a given coarse-graining scale,
we have introduced statistical homogeneity measures associated with particular quasi-local properties of quantum spacetime. 
By construction, these measures are very nonlocal and characterize the underlying quantum geometry in its entirety, through the statistics of 
the volume-averaged properties associated with a geodesic ball covering. This is reminiscent of the nonlocal
character of the global symmetry transformations that are used to characterize a smooth classical geometry as homogeneous and isotropic.  

Our application of the relative quantum homogeneity measures to two-dimen\-sional quantum gravity
led to interesting and apparently sensible results, demonstrating the viability of this new class of nonperturbative quantum observables.
It also highlighted the dependence of the homogeneity or otherwise on the local property under consideration, and the importance of
distinguishing between properties with a well-defined continuum interpretation and general discretization-dependent lattice properties. 
We do not claim uniqueness of our construction, and some of its elements can clearly be varied.  
The choice of how a geometry is covered by or divided into averaging regions is one example, although we do not expect it to
influence results substantially, as long as one
makes sure that the prescription is truly geometric and does not introduce an unwanted coordinate dependence or other biases.   

However, there is another and more subtle issue that could influence the outcome of our homogeneity analysis, namely the choice of
what we mean by ``averaging". For good reasons, which were explained in Sec.\ \ref{sec:homo} above, 
we have presently confined ourselves to studying local or quasi-local {\it scalar} quantities $Q(x)$. Although it is natural and simple to
construct a coarse-grained quantity $Q(x,\delta)$ by taking a volume average of $Q(x)$ over a geodesic ball (or some other local region of linear size
$\delta$), as we did in eq.\ (\ref{olocal}), this is not necessarily a unique or the physically most appropriate choice. This has to do with the complex
nature of local geometry, which in a continuum setting is captured by the tensorial character of the metric and curvature.
For instance, a scalar may have a complicated, composite dependence on local curvature tensors and their covariant derivatives, as long as we make sure
that all indices are contracted. However, an averaging of the individual components of a scalar $Q(x)$, however we define it, may not commute with 
the composition of its various components, leading to an ambiguity in what we may mean by ``the averaged scalar $Q(x,\delta)$".    

Related issues appear also for more general, nonsmooth geometries. For instance, the way in which
we determined the quasi-local Hausdorff dimension $d_h(x)$ from the metric properties of a piecewise flat geometry in Sec.\ \ref{subsec:haus} 
involved a rather complex operation of fitting to sphere volumes centred at $x$, according to eq.\ (\ref{sfit}).
One way to determine an averaged $d_h(x,\delta)$ is to average
$d_h(x)$ over the $\delta$-ball, as we did above, but one could also first average the sphere volumes over the $\delta$-ball and then extract an averaged
Hausdorff dimension by fitting to these averaged volumes, which turns out to lead to a different result, with larger inhomogeneity. 
Whether a given way of averaging is
appropriate is ultimately a question of how one operationally probes physics at the scale $\delta$, an issue that goes far beyond the
scope of this paper. The only points we want to highlight here is
that sticking to scalars is not necessarily a panacea, and that the way in which we implement averaging
can in principle affect the outcomes of quantum homogeneity measures.          

As already discussed in the introduction, a primary motivation for the present work and an
important application is to assess the homogeneity properties of the de Sitter-like ground state of
CDT quantum gravity in four dimensions. To extract reliable data in a sufficiently large $\delta$-interval will present 
computational challenges and may only be feasible for selected local properties.   
Whether approximately homogeneous behaviour is present and whether its onset can be observed in a regime that is accessible 
computationally will have to be investigated. 
An application of the quantum homogeneity measures to other formulations of quantum gravity and other ``quantum spacetimes",
obtained dynamically or otherwise, should in principle be possible, given that our construction only relies on elementary metric ingredients.   
Of course, any actual implementation will hinge on the ability to perform (numerical) computations in a near-Planckian regime and 
the availability of well-defined local quantities whose quantum homogeneity measures can be evaluated. 

\vspace{0.6cm}
\noindent {\bf Acknowledgement.} We thank the Albert Einstein Institute (Max-Planck Institute for Gravitational Physics) in Potsdam for hospitality in late 2022, where part of this work was done.

\section*{Appendix}

In Sec.\ \ref{sec:homo}, we commented on the difference between our homogeneity measures and those previously 
introduced in \cite{Cooperman2014} and applied there to CDT quantum gravity in three dimensions. The primary difference is
the presence of an explicit averaging (over geodesic balls) and of an associated averaging scale (the ball radius $\delta$) in our set-up.
This averaging scale is distinct from the scale intrinsically associated with a geometric quantity whose variance and thus homogeneity is 
considered, the scale we have called $r$ in the main body of the paper. To illustrate qualitatively that the $\delta$- and $r$-dependence of a homogeneity
measure are in general different, we will consider the homogeneity of the quantity $Q(x;r)\! =\! {\rm vol}({\cal S}_x(r))$, where the corresponding
averaged quantity $Q(x,\delta ;r)$ is obtained by averaging the shell volumes ${\rm vol}({\cal S}_x(r))$ over $\delta$-balls
according to the prescription (\ref{olocal}). This is similar to the volumetric homogeneity measure in \cite{Cooperman2014},
which was based on the volume of geodesic balls $B_x(r)$ rather than the volume of geodesic shells.\footnote{In our notation, 
the volumetric measure of \cite{Cooperman2014} corresponds to (the square of) our homogeneity measure ${\cal H}_Q (\delta\! =\! 0)$, 
where $Q(x;r)\! =\! {\rm vol}(B_x(r))$.}

\begin{figure}[t]
\centerline{\scalebox{0.48}{\rotatebox{0}{\includegraphics{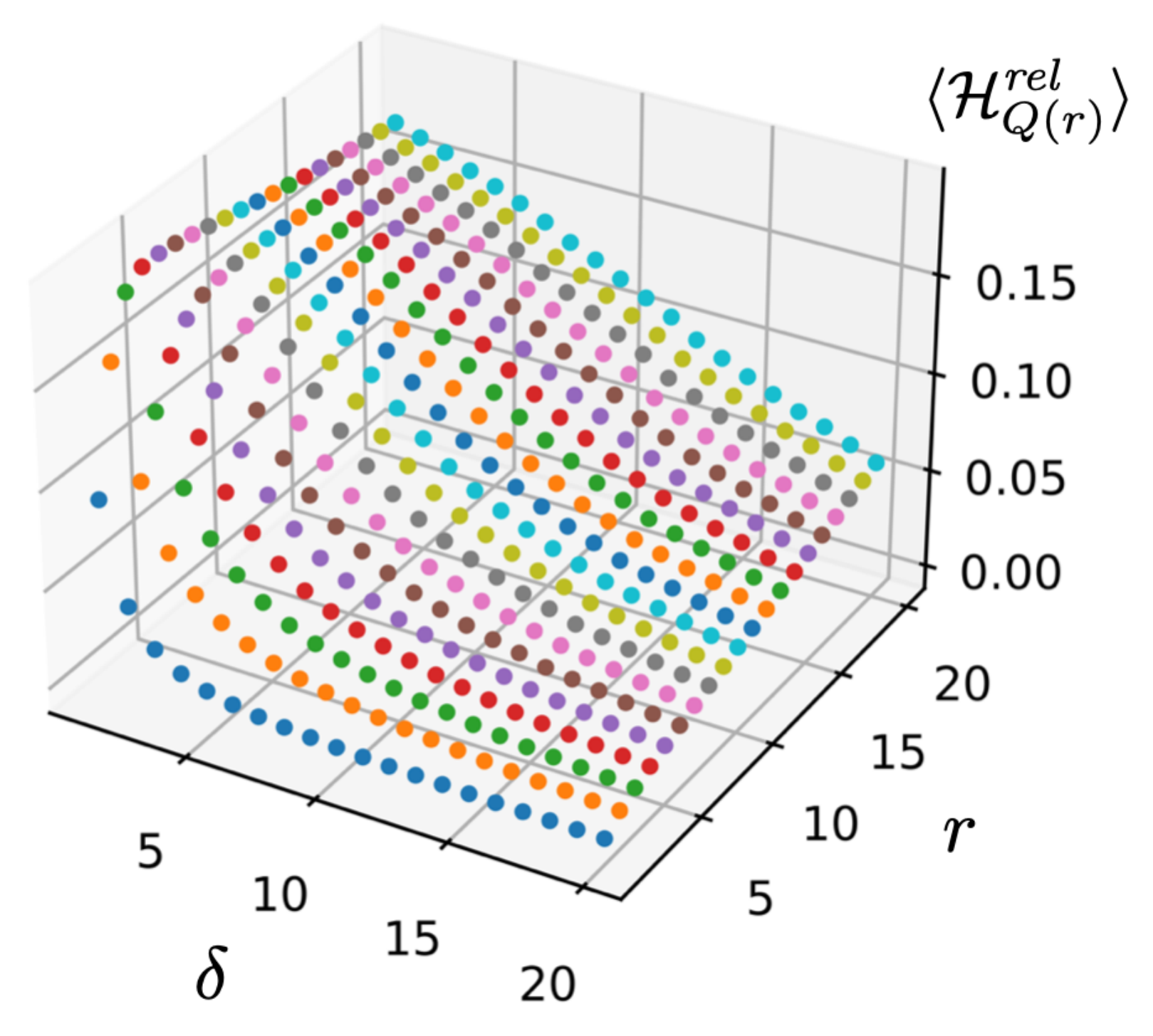}}}}
\caption{Expectation value  $\langle {\cal H}^{rel}_{Q(r)}(\delta)\rangle$ of the relative homogeneity measure for $Q(r)\! =\! {\rm vol}({\cal S}(r))$,
the volume of spheres of radius $r$, as a function of $r$ and the averaging scale $\delta$. Measurements taken on an
ensemble of two-dimensional CDT configurations of volume $N_2\! =\! 80k$ and time extension $t_{\rm tot}\! =\! 82$.}
\label{fig:rdelta}
\end{figure}

Fig.\ \ref{fig:rdelta} shows the expectation value $\langle {\cal H}^{rel}_{Q(r)}(\delta)\rangle$ of the relative homogeneity measure for 
$Q(r)\! =\! {\rm vol}({\cal S}(r))$, evaluated on two-dimensional CDT geometries with $N_2\! =\! 80k$ and $t_{\rm tot}\! =\! 82$,
based on data from 500 sampled configurations $T$. The range for both parameters is $[1,20]$. 
The data points for fixed $r\! =\! 1$ (the front sequence of dark blue dots in Fig.\ \ref{fig:rdelta}) coincide with those of the relative homogeneity 
measure for the coordination number of Sec.\ \ref{subsec:coord}, since the discrete length of a spherical shell of radius 1 at a vertex $x$ is always
equal to the coordination number of $x$. They display the familiar rapid fall-off as a function of the averaging scale $\delta$.
Examining other sequences of data points at fixed $r$, they are also monotonically decreasing as a function of $\delta$, but their decline becomes 
slower with increasing $r$. Note that from the perspective on averaging taken in the present work, one would usually consider only the region
$\delta\! \gtrsim\! r$, such that the averaging scales $\delta$ are larger than the linear scale $r$ characterizing the quantity that is being averaged.    
What comes closest to the situation studied in \cite{Cooperman2014} in our case are the data points for fixed minimal averaging scale 
(here given by $\delta\! =\! 1$) as a function of $r$. Even if one disregards the initial region $r\!\leq\! 5$ to avoid discretization artefacts,
$\langle {\cal H}^{rel}_{Q(r)}(\delta\! =\! 1)\rangle$ as a function of $r$ behaves very differently from $\langle {\cal H}^{rel}_{Q(r\! =\! 1)}(\delta)\rangle$
as a function of $\delta$, as
we claimed above. If one takes the former as a measure of relative homogeneity, it decreases beyond $r\! =\! 4$, but at a 
rather slow rate. At $r\! =\! 20$ it is still above 15\% and its behaviour is therefore not par\-ti\-cularly homogeneous in the range we have investigated.

\end{document}